\pdfoutput=1
\documentclass{raa_rb}
\usepackage{graphicx,times}
\usepackage{natbib}
\usepackage{amssymb,amsmath}

\usepackage{subfigure}
\usepackage{url}
\usepackage{CJK}

\renewcommand{\vec}[1]{ {\mathbf #1} }

\newcommand{\Fig}{{Figure}}
\newcommand{\Figs}{{Figures}}

\newcommand{\divB}{\nabla\cdot\mathbf{B}}
\newcommand{\crlB}{\nabla\times\mathbf{B}}

\bibpunct{(}{)}{;}{a}{}{,}

\begin{document}

\title{Testing a Solar Coronal Magnetic Field Extrapolation Code with
  the Titov--D{\'e}moulin Magnetic Flux Rope Model}

\author{Chaowei Jiang,
        Xueshang Feng}

   \institute{ SIGMA Weather Group, State Key Laboratory for Space
  Weather, Center for Space Science and Applied Research, Chinese
  Academy of Sciences, Beijing 100190, China; {\it cwjiang@spaceweather.ac.cn}\\
\vs \no
   {\small accepted by RAA}
}

\abstract{In the solar corona, magnetic flux rope is believed to be a
  fundamental structure accounts for magnetic free energy
  storage and solar eruptions. Up to the present, the extrapolation of
  magnetic field from boundary data is the primary way to
  obtain fully three-dimensional magnetic information of the
  corona. As a result, the ability of reliable recovering coronal
  magnetic flux rope is important for coronal field
  extrapolation. In this paper, our coronal field extrapolation code
  \citep[CESE--MHD--NLFFF,][]{Jiang2012apj} is examined with an
  analytical magnetic flux rope model proposed by
  \citet{Titov1999}, which consists of a bipolar magnetic configuration
  holding an semi-circular line-tied flux rope in force-free equilibrium.
  By using only the vector field in the bottom boundary as input,
  we test our code with the model in a representative range of parameter space
  and find that the model field is reconstructed with high accuracy.
  Especially, the magnetic topological interfaces formed between the flux rope and the surrounding
  arcade, i.e., the ``hyperbolic flux tube'' and ``bald patch separatrix
  surface'', are also reliably reproduced. By this test, we demonstrate that
  our CESE--MHD--NLFFF code can be applied to recovering magnetic flux
  rope in the solar corona as long as the vector magnetogram satisfies
  the force-free constraints.
}

   \authorrunning{C.-W. Jiang et al. }            
   \titlerunning{Testing NLFFF Extrapolation Code with Flux Rope Model}  
   \maketitle

%


\section{Introduction}

The magnetic field plays a fundamental role in all physical processes
in the Sun's corona, such as the formation of coronal loops and
prominences (or filaments), the production of solar flares, filament
eruptions, and coronal mass ejections, as well as the determination of the
structure of solar wind \citep{Solanki2006}. However, it is very difficult to
make a direct measurement of the coronal magnetic field. Works that have been done to measure the coronal fields using the radio and infrared wave bands \citep[e.g.,][]{Gary1994,Lin2004}
can only give fragmentary and occasional data. Up to the present, the routine measurement
of the Sun's magnetic field that we can rely on is restricted to the solar surface, i.e., the photosphere. This is
extremely unfortunate since the role of the magnetic field playing in
the corona is much more important than that in the photosphere. As a result,
our knowledge of the three-dimensional (3D) coronal magnetic
field is largely based on extrapolations from photospheric
magnetograms using some kind of reasonable physical models. In the low corona where the plasma
$\beta$ (the ratio of gas pressure to magnetic pressure) is rather
small $(\sim 0.01)$, the magnetic field can be well assumed as free of
Lorentz force in the case of quasi-static state (i.e., $\vec J\times
\vec B=\vec 0$ where $\vec J=\crlB$ is the current and $\vec B$ is
the magnetic field). Thus the force-free field model is usually adopted in
coronal field extrapolations.

Owing to the difficulty of direct solving the force-free equation
$(\crlB)\times\vec B=\vec 0$ which is intrinsically nonlinear,
a variety of numerical
codes have been proposed for nonlinear
force-free field (NLFFF) extrapolations (e.g., see review papers by \citet{Schrijver2006},
\citet{Metcalf2008}, \citet{Wiegelmann2008}).  For faster convergence and better accuracy over the
available codes, the authors have developed a new extrapolation
code called CESE--MHD--NLFFF \citep{Jiang2011,Jiang2012apj}, which is
based on magnetohydrodynamics (MHD) relaxation method and an advanced
numerical scheme, the spacetime conservation-element/solution-element
(CESE) method. We have also critically examined our code with several NLFFF
benchmark models and compared the results with previous
joint studies by \citet{Schrijver2006} and \citet{Metcalf2008}, which demonstrates
its performance. The code has also been extended to application in
spherical geometry and seamless full-sphere extrapolation for the
global corona \citep{Jiang2012apj1}.

Coronal magnetic flux rope (MFR) is of great interest in the study of
solar eruptive activities like filament eruption and CMEs. It is
believed to be a good candidate for the critical pre-eruptive
structures of storing magnetic free energy and helicity and holding
cold dense filament material against gravity, while its instabilities
can account for triggering and driving of eruptions.  Observationally,
a sequence of evidences such as the coronal sigmoid \citep{Rust1996,
  Canfield1999}, coronal hot channels \citep{ZhangJ2012, ChengX2013},
and coronal cavities \citep{Gibson2006, Regnier2011} all suggested the
existence of pre-eruptive MFR.  Theoretically, MFR is an essential
building block of many flare and CME models
\citep[e.g.,][]{Forbes1991, Chen2000, Wu2000, Torok2005, Kliem2006}.
In the coronal field extrapolations, it has been reported frequently
that MFRs consistent with observations are reconstructed from relevant
photospheric magnetograms \citep[e.g.,][]{Canou2010, Cheng2010,
  Guo2010, Jing2010, GuoY2013, Jiang2014formation, Jiang2014NLFFF}.
Although many NLFFF codes has been demonstrated with the ability of
extrapolating MFR, the reliability is still not fully approved. Thus,
in the paper, we examine the reliability of our CESE--MHD--NLFFF code
for extrapolating coronal MFR using an analytic force-free model
proposed by \citet[][hereafter TD model]{Titov1999}.  The remainder of
the paper is organized as follows. Section~\ref{sec:code} describes
briefly the CESE--MHD--NLFFF code. The TD model is described in
Section~\ref{sec:TD}.  Extrapolated result are shown in
Section~\ref{sec:res} and our conclusions are summarized in
Section~\ref{sec:con}.

\section{The CESE--MHD--NLFFF Code}
\label{sec:code}

In using the MHD relaxation approach for achieving a NLFFF, one
usually starts from a potential field model matching the vertical
component of the magnetogram, then replaces the transverse fields on
bottom boundary with those from vector magnetogram (which is obviously
inconsistent with the potential value, thus drives the system to
evolve dynamically), and finally lets the MHD system to seek a new
equilibrium in which all the other forces are negligible if compared with
the Lorentz force. Consequently the Lorentz force must be nearly
self-balancing and the final field can be regarded as the target solution of
magnetic force-freeness. In our CESE--MHD--NLFFF code, we solve a simplified
zero-$\beta$ MHD model with a fictitious frictional force, which is
used to assure that a final equilibrium can be reached in a smooth way
\citep{Roumeliotis1996, Valori2007}.  The specific equation is written in the
following form with magnetic splitting
\begin{eqnarray}
  \label{eq:main_equ}
  \frac{\partial\rho\mathbf{v}}{\partial t} =
  (\crlB_{1})\times\mathbf{B}-(\divB_{1})\vec B-\nu\rho\mathbf{v},
  \nonumber\\
  \frac{\partial\mathbf{B_{1}}}{\partial t} =
  \nabla\times(\mathbf{v}\times\mathbf{B})
  +\nabla(\mu\divB_{1})
  -\mathbf{v}\divB_{1},
  \nonumber\\
  \rho=|\vec B|^{2},\ \
  \mathbf{B}=\mathbf{B}_{0}+\mathbf{B}_{1}
\end{eqnarray}
where the total magnetic field $\vec B$ is split into
$\vec B_{0}$, a potential field matching the normal component of the magnetogram,
and $\vec B_{1}$, the deviation between $\vec B$ and $\vec
B_{0}$. The last two terms in the induction equation are used to
suppress the numerical errors of magnetic divergence (i.e.,
numerically induced magnetic monopole). $\nu$ is the frictional
coefficient and $\mu$ is the diffusive speed of the numerical magnetic
monopole. The value of them are  given by $\nu=1/(5\Delta
t)$ and $\mu=0.4(\Delta x)^{2}/\Delta t$ in the code, respectively, where $\Delta t$
 is the time step and $\Delta x$ is the grid size. More details and the advantages
of using above equations can be found in
\citep{Jiang2012apj,Jiang2012apj1}.

As the above equation system is just a simplified subset
of the full MHD system, any available MHD
code can be used to solve it. By taking into account the computational
efficiency and accuracy, we prefer to utilize modern codes for MHD.
However, most of the modern MHD codes are based on theory of
characteristic decomposition of a hyperbolic system, and they are not
suitable for Equation~\ref{eq:main_equ}, because it is not a hyperbolic system.
We thus select the CESE--MHD scheme
\citep{Jiang2010}, which is free of characteristic decomposition and
is very suitable for Equation~\ref{eq:main_equ}. Furthermore, the CESE--MHD
code has been successfully applied in solving many relevant problems in solar physics,
e.g., the dynamic evolution of AR using data-driven MHD model \citep{Jiang2012c},
the global corona structure
\citep{Feng2012apj} and the interplanetary solar wind modeling
\citep{Feng2012SoPh,Yang2012JGR}.

\begin{figure*}[htbp]
  \centering
  \includegraphics[width=0.8\textwidth]{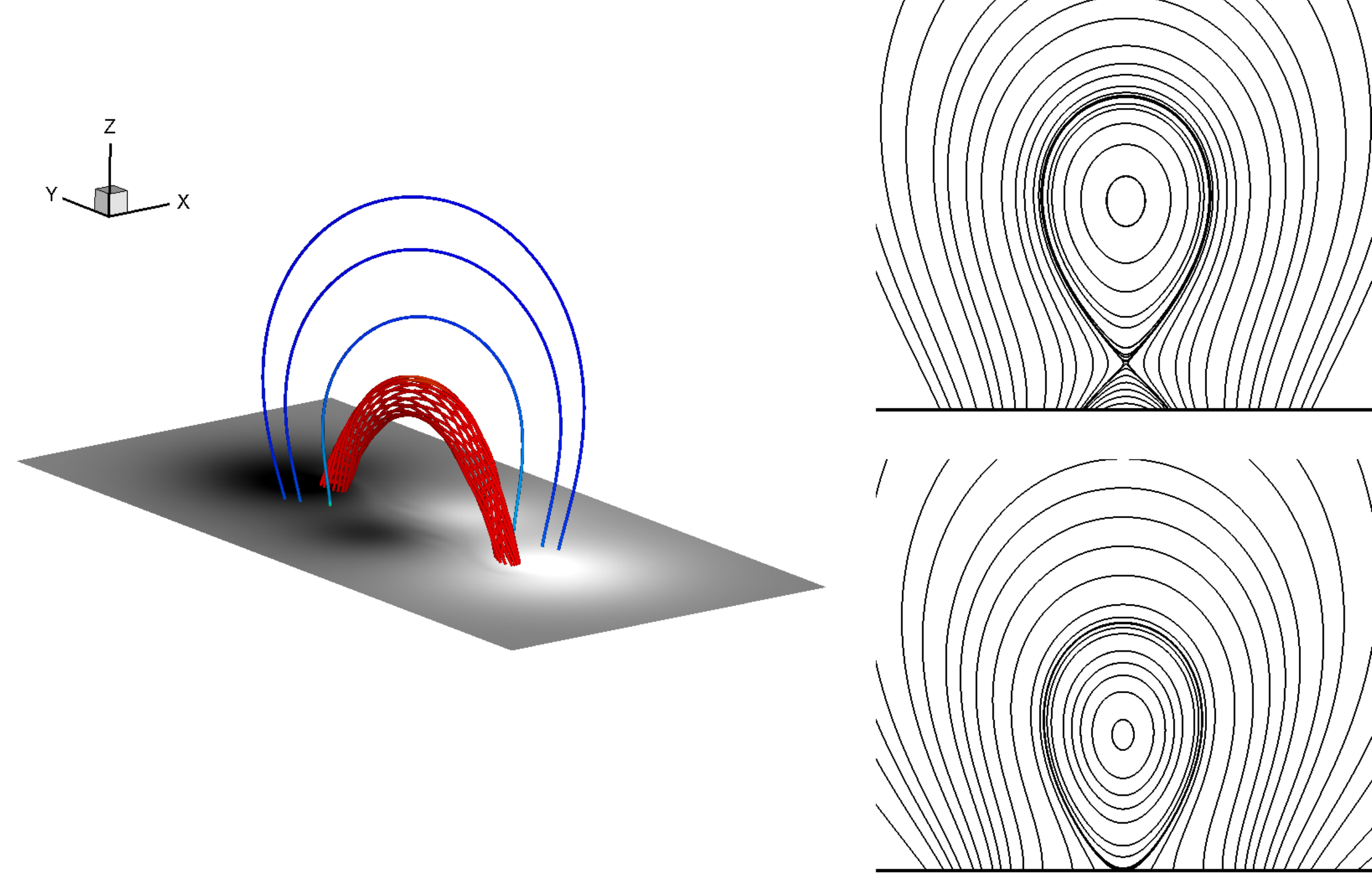}
  \caption{Left: basic magnetic configuration of TD model, the flux
    rope is shown with the red lines, the overlying near-potential
    arcades with blue lines, and the grey image shows the normal
    magnetogram on the bottom.  Right: central vertical cross section
    of TD models with HFT (top, the 2D field lines form a X-point up
    in the corona) and with BP (bottom, the 2D field lines form a
    tangency point with the photosphere), illustrating two different
    stages of an emerging flux rope, fully emerged and partial
    emerged, respectively.}
  \label{fig:sample}
\end{figure*}

\section{The TD Flux Rope Model}
\label{sec:TD}

The TD flux rope model has gained considerable interest because of its relevance
to the structures of solar active regions and eruptive magnetic field, which is
demonstrated by many investigations \citep{Roussev2003, Torok2004,
  Torok2005, Schrijver2008, McKenzie2008, Titov2014}. Basically, the
TD model is constructed to simulate an bipolar active region
containing a half-buried, toroidal current-carrying flux rope balanced
(or confined) by the overlying arched near-potential field (as shown in
\Fig~\ref{fig:sample}).  Excessive twist of the flux rope can trigger
its kink instability \citep{Torok2005}, and a too fast decaying of the
overlying field with height can trigger the torus instability of the
system \citep{Kliem2006}.
With the change of its parameters, the model can also be used to
illustrate different stages of an twisted sub-photospheric flux tube
emerging {\it bodily} into the corona \citep[see Figure~7
of][]{Gibson2006}. As shown in \Fig~\ref{fig:sample}, two different
configurations, the one with a bald patch separatrix surface (BPSS), and
the one without BPSS but with a hyperbolic flux tube (HFT), show the
stages of partial and full emergence of flux rope, respectively. The
reader is referred to \citep{Titov1999, Titov2014, Valori2010} for
detailed description of the model and its parameter settings.

By only using the magnetic field on the model's bottom boundary to
reconstruct the flux rope, the TD model represents a far more
difficult challenge of extrapolation than those simple sheared field models
\citep[e.g., the force-free field model by][]{Low1990}.
As examined by \citet{Wiegelmann2006TD}
and \citet{Valori2010}, this model requires
a topological change from the initial potential field to obtain the flux-rope configuration.
\citet{Valori2010} has extensively tested their
extrapolation code using the TD model with a series of parameter sets,
which includes four sets of stable model. They are, respectively, a Low-HFT case, a High-HFT
case, a No-HFT case and a BP case. An HFT is present in the first two cases, one
with the HFT very close to the photosphere (Low-HFT) and the other
with the HFT reaching significantly into the volume (High HFT).
The No-HFT case has no HFT present above the photosphere, hence, its
magnetic topology is much simpler than the first two cases. These three
configurations have no bald patch at the photosphere because
the toroidal field component is relatively strong. For the BP case, the flux rope has a left-handed average twist of about $2\pi$, which is close to the twist of the first three equilibria, but a BPSS is introduced in the resulting field by enlarging the minor radius of the torus. Here we use exactly the
same reference data from \citet{Valori2010}'s paper with the same grid resolution $\Delta=0.06$ and
extrapolation box of interest, $[-3.03, 3.03]\times [-4.95, 4.95]\times [-0.06, 4.44]$,
while our actual computational volume is several times larger than the extrapolation volume of interest
to minimize the numerical boundary effects.

\begin{figure}[htbp]
  \centering
  \includegraphics[width=0.6\textwidth]{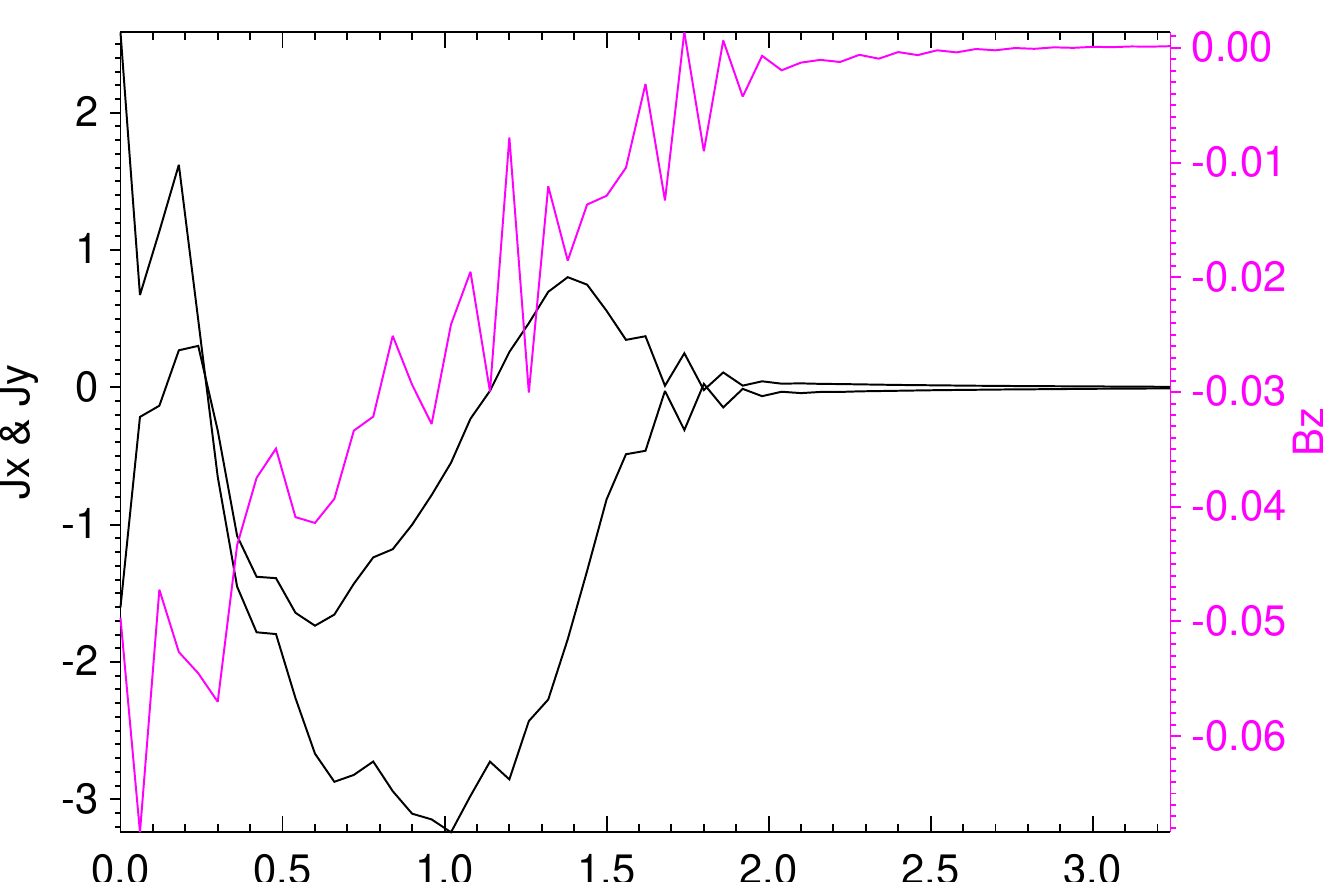}
  \caption{Numerical oscillation of the reference data of the TD
    model: $J_x, J_y$ and $B_z$ along the vertical grid line of $x=-0.03,
    y=-0.03$ in the High\_HFT case.}
  \label{fig:osc}
\end{figure}

Because the analytical TD solutions are approximately force-free, \citet{Valori2010}
relaxed them to numerical equilibria using the MHD code of
\citet{Torok2003}, which results in only little change to the geometry shape of
the flux rope but improves the force-freeness for the models. Even though, we should point out that these
reference models are still not perfect force-free solutions for the
following reasons.  Firstly, the data contains strong numerical
oscillation, for example, see \Fig~\ref{fig:osc}. Although not obvious
in tracing the field lines which are integral result of the field
data, the oscillation is obvious when make numerical difference on the
data.  This oscillation is resulted from the MHD relaxation of the
analytic TD model by the \citet{Torok2003}'s code. Secondly, the
bottom magnetograms (i.e., the vector field on the bottom boundary of
model data) contains force that cannot be ignored. To assess the
force-free quality of the magnetograms, we calculated the same metrics
$\epsilon_{\rm flux}$, $\epsilon_{\rm force}$ and $\epsilon_{\rm
  torque}$ as in \citep{Wiegelmann2006, Jiang2013NLFFF}, which measure
flux, force, and torque imbalance of the magnetogram.
As can be seen, the force-freeness is fulfilled well by the first
three cases, but not that well for the BP case (note that both
parameters $\epsilon_{\rm force}$ and $\epsilon_{\rm torque}$ is above
0.01), which can cause non-negligible inconsistence in the NLFFF
extrapolation. For these reasons, a perfect
extrapolation does not necessarily mean to
reproduce a magnetic field matching perfectly the reference model.

\begin{table}[htbp]
  \centering
  \caption{Quality of the magnetograms: normalized flux $\epsilon_{\rm flux}$,
    force $\epsilon_{\rm force}$, and torque $\epsilon_{\rm torque}$
    imbalance as defined in \citep{Wiegelmann2006}.}
  \begin{tabular}{lrrr}
    \hline
    \hline
    Case & $\epsilon_{\rm flux}$ & $\epsilon_{\rm force}$ & $\epsilon_{\rm torque}$ \\
    \hline
    High\_HFT & 1.30E-08 & 3.45E-03 & 5.48E-03 \\
    Low\_HFT  & -3.63E-08 & 5.88E-03 & 8.88E-03  \\
    No\_HFT   & 5.83E-09  & 7.29E-03 & 1.09E-02 \\
    BP       & 2.08E-08  & 1.46E-02 & 2.09E-02\\
    \hline
 \end{tabular}

 \label{tab:quality}
\end{table}

\section{Results}
\label{sec:res}

Same as \citet{Valori2010}, the extrapolation results are analyzed
within a central region of the extrapolation box by discarding
20 grid layers of its top and lateral boundaries. The metrics for accessing
the extrapolation quality are defined as \citet{Jiang2013NLFFF}. Similarly,
we measure the
degree of force-freeness by the current-weighted (and
current-square-weighted) average sine-angle between magnetic field and current,
\begin{equation}
  {\rm CWsin} \equiv
  \frac{\int_{V}{J}\sigma dV}{\int_{V}{J} dV};
  \ \
  {\rm C^{2}Wsin} \equiv
  \frac{\int_{V}J^{2}\sigma dV}{\int_{V}J^{2} dV},
  \ \
  \sigma =  \frac{|{\vec J}\times{\vec B}|}{{J}{B}}.
\end{equation}
The solenoidal property is quantified by
\begin{equation}
  \langle |f_{i}|\rangle =
  \frac{1}{V}\int_{V}\frac{\divB}{6B/\Delta x} dV.
\end{equation}
The mean relative error between the extrapolated field $\vec B$ and
the original reference field $\vec B_{\rm ref}$,
\begin{equation}
  E_{\rm M} = \frac{1}{V}\int_{V} \frac{|\vec B^{\rm ref}-\vec
    B|}{|\vec B^{\rm ref}|}.
\end{equation}
In order to reliably compare the reference and the extrapolated fields
and their qualities of force-freeness, we use a fourth-order
difference of $\vec B$ to calculate the current $\vec J$ and
$\divB$. Besides, to judge the magnetic topology, we compute the
relative errors of the apex heights of the flux rope axis (FRA) and
the HFT (if present) between the extrapolated and reference fields.
Since the flux rope writhes only slightly out of
the plane ${x = 0}$ in the models considered in this paper, both of the
apex heights can be approximately as the inversion points of $B_x(0, 0, z)$,
which is a good approximation of the poloidal
component of the field at the line-symmetric $z$ axis.
Finally we compute the relative errors of the magnetic
energy ($E_{\rm mag}$) between the extrapolated and reference fields.

The results are given in Table~\ref{tab:metrics}. Regarding the
metrics of force-freeness and divergence-freeness, the extrapolation
code achieves solutions slightly even better than the reference models
for the first three cases. Since both the models (i.e., the reference
model and the extrapolation) are produced by numerical codes, it is
suggested that our code performs better in relaxing the magnetic field
to a force-free and divergence-free solution.  For the first three
cases, the mean relative errors $E_{\rm m}$ are only several percents,
demonstrating that the reference models are reproduced with very high
accuracy. The topology parameters, e.g., apex of FRA and HFT are very
close to those of the reference model, except for the HFT apex of the
low HFT case with a relative error of fifteen percent. This is because
in the low HFT case the HFT is rather low with only about one grid
point from the bottom, thus the size of the grid is not sufficiently
small to resolve the HFT.  The force-freeness for the BP case is not
as good as the first three cases, and consequently the topology of the
extrapolated field also deviates considerably from the reference
model. This is as expected because we have shown that the magnetogram
of the BP case is most inconsistent with the force-free constraints
(Table~\ref{tab:quality}). For all the cases, the energy content is
well recovered with relative errors below several percents.

\begin{table*}[h!]
  \centering
  \caption{Results for the metrics of extrapolations}
  \begin{tabular}{crrrr}
    \hline
    \hline
    Figure of merit             & High\_HFT & Low\_HFT & No\_HFT & BP \\
    \hline
    CWsin $\times 10^{2}$        & 2.07/2.59  &  2.04/2.34 & 2.25/1.94 &5.41/0.81 \\
    C$^{2}$Wsin $\times 10^{2}$  &  1.04/1.24  &  0.82/1.08 & 0.77/0.91 & 2.50/0.41  \\
    $<|f_{i}|> \times 10^{5}$    &  4.06/6.52  &  3.28/6.45 & 3.52/5.74& 12.7/7.57   \\
    $E_{\rm M}$                   & 0.020       & 0.015     &  0.016 &   0.116        \\
    HFT apex                     & 5.18\%      & 15.5\%    &   ...   &  ...   \\
    FRA apex                     & 5.03\%      & 1.38\%    &  0.05\% &  22.1\%\\
    $E_{\rm mag}$                  & 0.75\%      & 0.98\%   & 1.1\% &     2.1\%         \\
    \hline
 \end{tabular}
 \tablecomments{0.86\textwidth}{For the first
   three metrics, results of the reference models are also given
   following those of the extrapolations.}
 \label{tab:metrics}
\end{table*}

\begin{figure*}[htbp]
  \centering
  \includegraphics[width=0.6\textwidth, angle=-90]{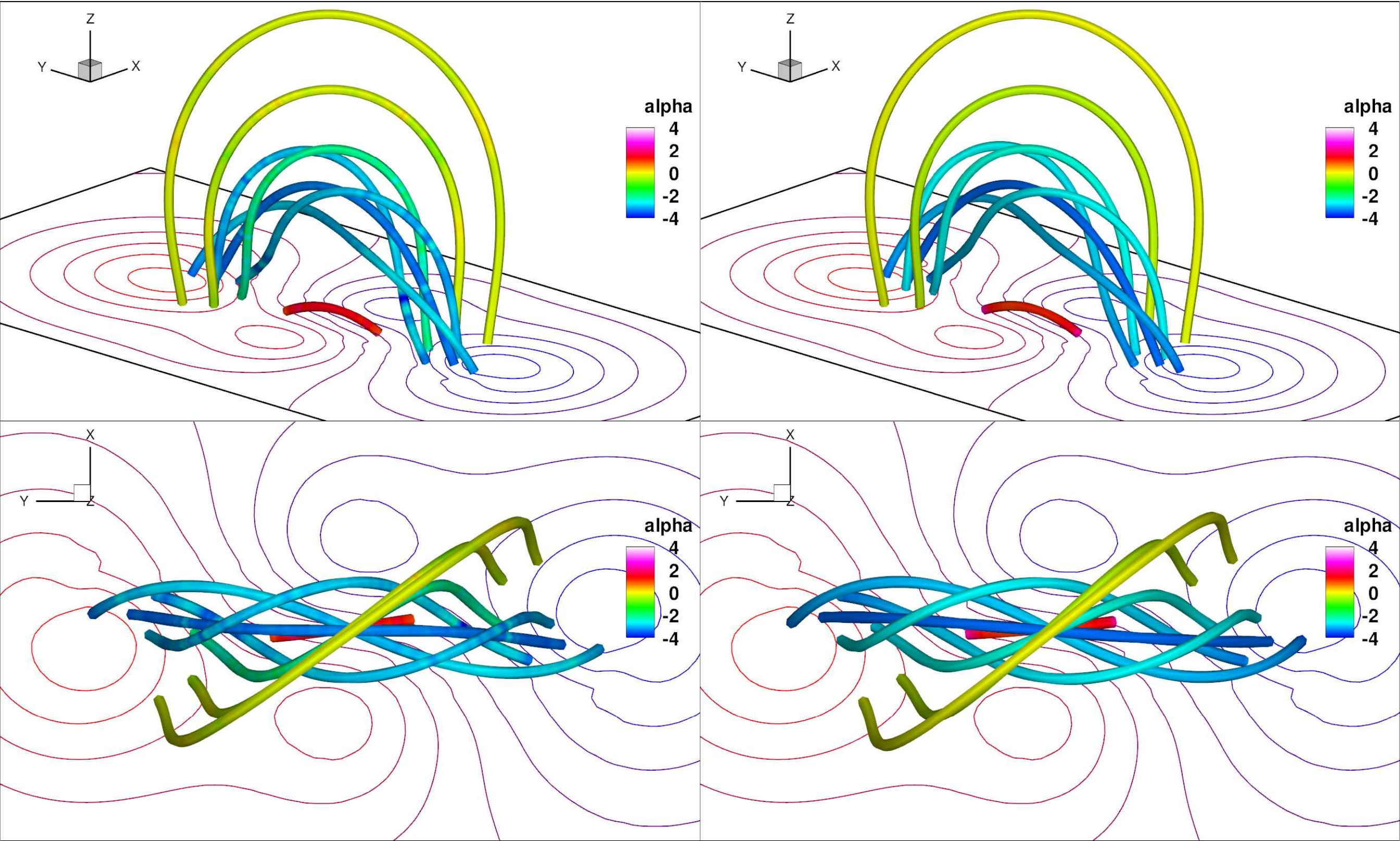}
  \caption{Selected field lines for The High\_HFT case: left are the
    reference model and right the extrapolation. Contour lines on the
    bottom are plotted for $B_z$. The field lines are color-coded by
    the value of force-free factor $\alpha$.}
  \label{fig:High_HFT}
\end{figure*}

\begin{figure*}[htbp]
  \centering
  \includegraphics[width=0.6\textwidth, angle=-90]{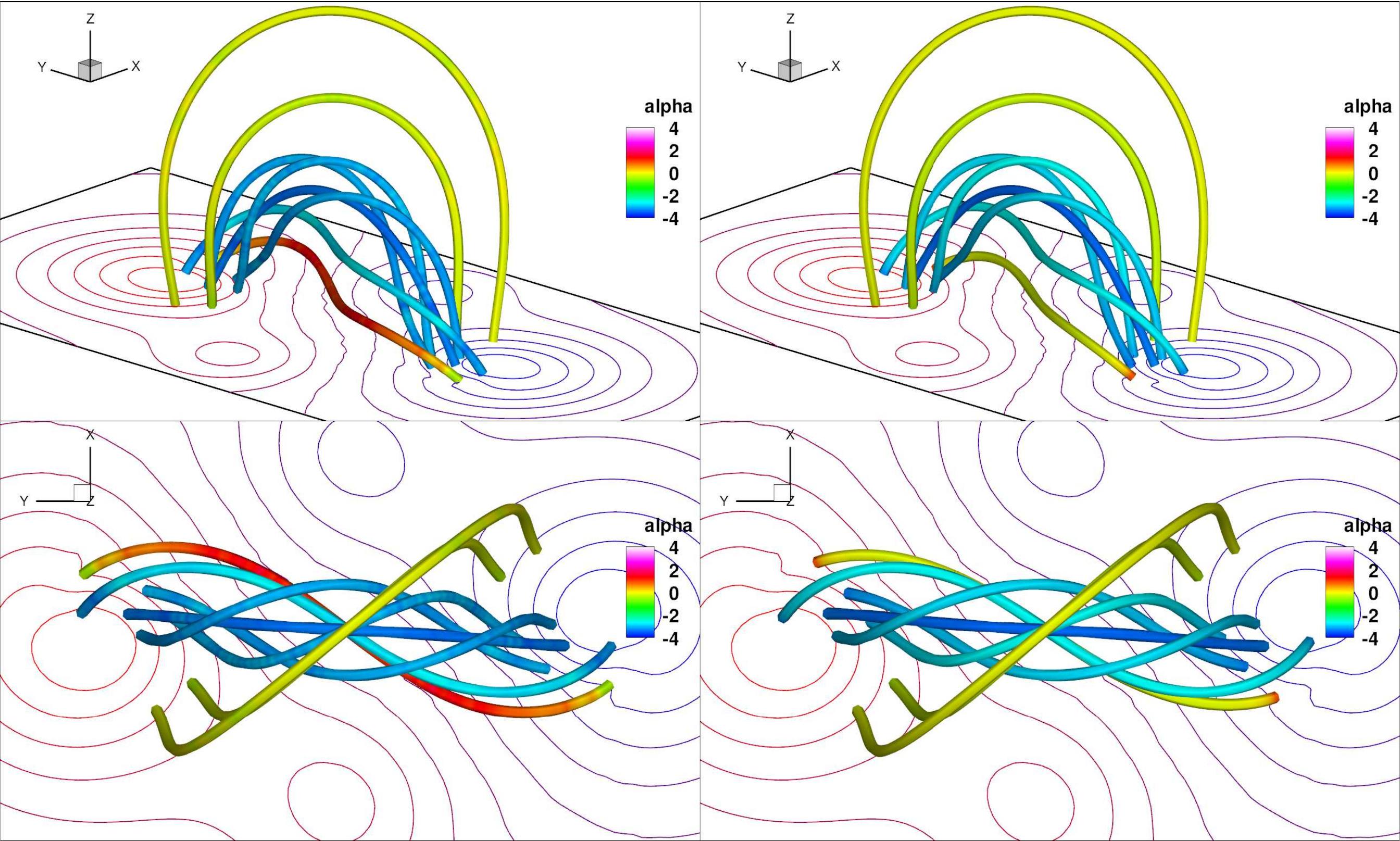}
  \caption{Selected field lines for The No\_HFT case. The format is
    the same as \Fig~\ref{fig:High_HFT}.}
  \label{fig:No_HFT}
\end{figure*}

\begin{figure*}[htbp]
  \centering
  \includegraphics[width=0.4\textwidth]{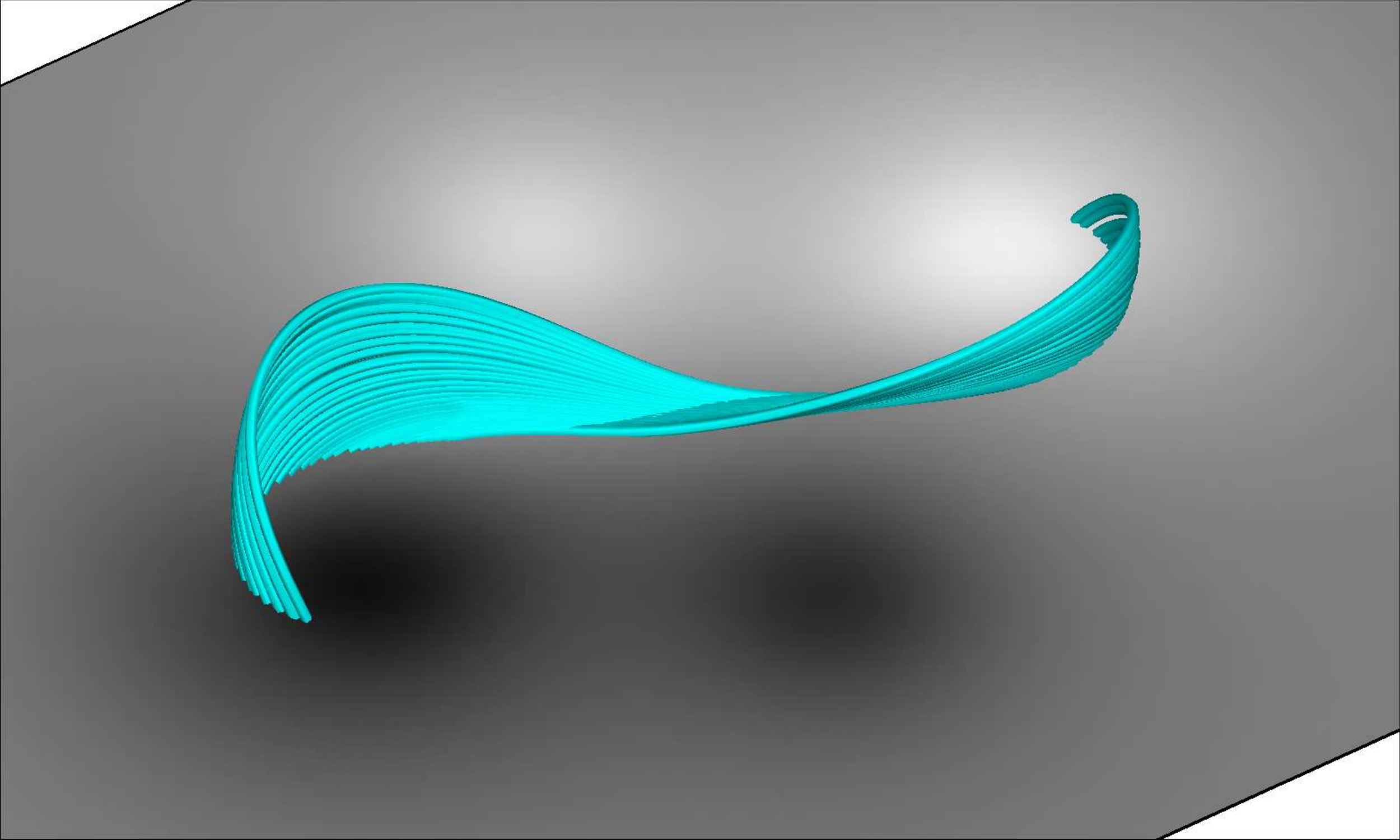}
  \includegraphics[width=0.4\textwidth]{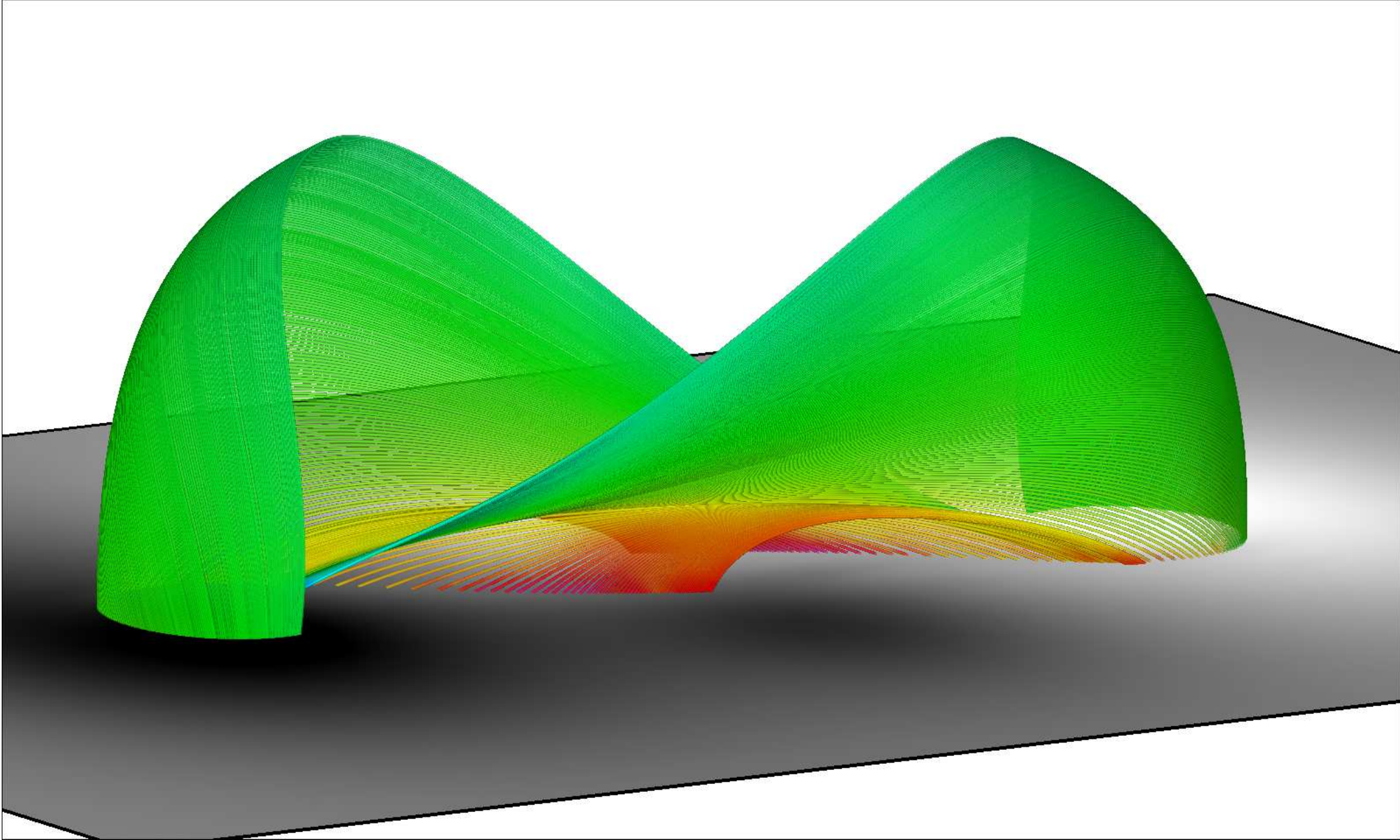}
  \caption{The reconstructed BPSS (left) and HFT (right) illustrated
    with continuous set of field lines (colors of the lines are used
    for a good visualization of the structures). They characterize the
    interface layers that separating the flux rope from its ambient
    field.}
  \label{fig:separatrix}
\end{figure*}

For a visual inspection of the magnetic configuration,
we show selected field lines of two cases, the High\_HFT and No\_HFT,
in \Figs~\ref{fig:High_HFT} and \ref{fig:No_HFT}, respectively.
The field lines for all models are traced from the same set of points
in the central cross section of the volume.
The field lines include the flux rope axis, four field lines
closely around the rope axis, one low-lying below the flux rope and two highly
overlying the flux rope. The field lines are color-coded by the value of force-free parameter $\alpha$.
The side-by-side comparison of geometry of the field lines
shows little difference, but the colors of the field lines differs.
Ideally for a force-free field, $\alpha$ should be constant along a given field line.
However, note that for the reference model, strong oscillation of the $\alpha$ can be seen on any
field line. In this respect, our extrapolation code gives a better
solution with the color much more uniform on any field line. Finally we show the
reconstructed BPSS and HFT in \Fig~\ref{fig:separatrix}.

\section{Conclusions}
\label{sec:con}

We have examined the CESE--MHD--NLFFF code by the TD flux rope
model. It is demonstrated that our NLFFF extrapolation code can
reconstruct flux ropes and their related topology structures (e.g.,
BPSS and HFT) reliably from only the bottom boundary data (i.e., the
vector magnetogram) from the model field. Although the extrapolation
is sensitive to the qualities of the vector magnetograms, the relative
errors with the reference field are rather small for all the test
cases in the paper. Basing on the present and all the previous tests
\citep[e.g., ][]{Jiang2012apj, Jiang2013NLFFF}, we are more confident
in applying our code to the realistic coronal field if the magnetogram
is preprocessed to fulfill the force-free constraints
\citep{Jiang2014Prep}.

\normalem
\begin{acknowledgements}

This work is jointly supported by the 973 program under grant
2012CB825601, the Chinese Academy of Sciences (KZZD-EW-01-4), the
National Natural Science Foundation of China (41204126, 41274192,
41031066, and 41074122), and the Specialized Research Fund for State
Key Laboratories. C.W.J is also supported by 
Youth Innovation Promotion Association of CAS (2015122). 
We are grateful to Dr. Valori G. for providing the
numerical data of the TD model.

\end{acknowledgements}


\end{document}